\documentclass[a4paper,fleqn,usenatbib]{mnras}

\usepackage{newtxtext,newtxmath}

\usepackage[T1]{fontenc}
\usepackage{ae,aecompl}

\usepackage{graphicx}	
\usepackage{amsmath}	
\usepackage{soul}

\title[Optical properties and dust temperatures in clumpy diffuse medium]{Optical properties and dust temperatures in clumpy diffuse medium}

\author[A. B. Ostrovskii et al.]{
Andrei B. Ostrovskii,$^{1}$\thanks{E-mail: Andrey.Ostrovsky@urfu.ru}
S. Yu. Parfenov,$^{1}$
A. I. Vasyunin,$^{1,\;2\thanks{Visiting Leading Researcher}}$
A. V. Ivlev,$^{3}$
\newauthor V. A. Sokolova$^{1,\;2}$
\\
$^{1}$Ural Federal University, Lenin av., 51, Yekaterinburg 620000, Russia\\
$^{2}$Engineering Research Institute <<Ventspils International Radio Astronomy Centre>> of Ventspils University of Applied Sciences, Inzenieru 101,\\ \;\ Ventspils, LV-3601, Latvia\\
$^{3}$Max-Planck-Institut f\"{u}r extraterrestrische Physik, D-85741 Garching, Germany
}

\date{Accepted XXX. Received YYY; in original form ZZZ}

\pubyear{2018}

\begin{document}
\label{firstpage}
\pagerange{\pageref{firstpage}--\pageref{lastpage}}
\maketitle

\begin{abstract}
In this study, we explore the impact of inhomogeneities in the spatial distribution of interstellar dust on spatial scales of $\le1$~au caused by ion shadowing forces on the optical properties of diffuse interstellar medium (ISM) as well as on the dust temperature. We show that recently proposed possibility that interstellar dust grains in the diffuse ISM are grouped in spherical cloudlets (clumps) may significantly affect the observed optical properties of the diffuse ISM in comparison to that calculated under the commonly accepted assumption on the uniform dust/gas mixture if the size of clumps $\gtrsim0.1$~au. We found that opacity of an arbitrary region of diffuse ISM quickly decreases with growth of dusty clumps. We also studied the dependence of opacity and dust temperature inside the dusty clumps on their size. We show that the clumps larger than 0.1~au are opaque for far ultraviolet radiation. Dust temperature exhibit a gradient inside a clump, decreasing from the edge to the center by several degrees for a clump of a size of 0.1~au and larger. We argue that dust temperatures and high opacity within clumps larger than 0.1~au may facilitate somewhat more efficient synthesis of molecules on surfaces of interstellar grains in the diffuse ISM than it was anticipated previously. On the other hand, the presence of clumps with sizes below 0.1~au makes small or negligible influence on the optical properties of the diffuse ISM in comparison to the case with uniform dust/gas mixture.
\end{abstract}




\section{Introduction}

Interstellar medium (ISM) in the Galaxy is known to be inhomogeneous at various spatial scales. Average properties of the largest--scale inhomogeneity represented by different phases of the ISM are studied observationally and theoretically already for several decades, and, as of today, established relatively well~\citep[][]{Field_ea69, McKee_ea77, 2006ARAA..44..367S}. One of the ISM phases is diffuse interstellar medium, which is characterized by moderate temperatures ($\le 100$~K) and densities ($n_{\rm{H}}\le 10^{2}$~cm$^{-3}$), as well as low extinction against visible light and diffuse interstellar ultraviolet (UV) field ($A_V \le 1^m$)~\citep[][]{2006ARAA..44..367S}. Low gas densities and transparency for UV field makes the diffuse medium a hostile environment for interstellar molecules. Despite this, a number of molecules, including polyatomic, is found in the diffuse ISM to date~\citep[e.g.,][]{Thiel_ea17, Liszt_ea18, Gerin_ea19}. Observed abundances of many species in diffuse ISM reach relatively high values that cannot be explained without invoking assumptions on various small--scale energetic processes that presumably occur in the diffuse ISM~\citep[e.g.,][]{Gredel_ea93, Godard_ea12, Godard_ea14}. Some of the processes, in turn, can lead to the formation of small--scale inhomogeneities of the diffuse ISM. Indeed, there is growing observational evidence that density of diffuse gas can vary significantly on small spatial scales below 1~pc~\citep[see e.g.,][]{2018ARA&A..56..489S}.

It was shown theoretically that homogeneous dusty plasmas are intrinsically unstable \citep{2000PlPhR..26..682M,2001A&A...376L..43B,2003CoPP...43...51T}. The instability is triggered by attractive ion `shadowing' forces between grains, induced due to plasma absorption on the grain surfaces \citep{2008LNP...731..333T}. \citet{Tsytovich_ea14} showed that in astrophysical environments this leads to the formation of stable compact clumps of dust, with the grain number density orders of magnitude higher than in a homogeneous case.
Since virtually all dust particles are in the clumps, the interstellar gas between clumps is almost dust-free. As shown in~\citet[][]{Ivlev_ea18}, such a small--scale inhomogeneous dust distribution represented by dusty clumps may play an important role in chemical evolution of the diffuse ISM. In comparison to the `standard' case of uniform mixture of gas and dust, in the `clumpy' medium, formation of molecular hydrogen via surface processes may proceed up to an order of magnitude faster~\citep{Ivlev_ea18}. Obviously, possible presence of dusty clumps in the diffuse ISM must affect not only kinetics of the H~$\rightarrow$~H$_{2}$ transition, but also chemistry of other molecular species~--- a problem yet to be explored. Next, since interstellar dust particles are the principal agent that determines the opacity of the medium against visible and UV radiation, clumpy dust distribution must result in different optical properties of the diffuse ISM in comparison to the uniform gas/dust mixture. Different studies demonstrate that clumping decreases the effective medium opacity \citep[see][]{1999ApJ...523..265V, Conway_2018}. \citet{2015A&A...584A.108S} show that the observed extinction curve flatten in the clumpy medium with optically thick clumps. Quantitative evaluation of optical properties of the clumpy diffuse ISM is thus important for two reasons. First, it is needed for further studies of chemistry in diffuse ISM, as photochemical processes play a crucial role there. As noted by~\citet{1994ASPC...58..319V}, the molecules in diffuse regions and translucent clouds can dissociate in the medium between clumps and survive within the clumps. Indeed, the calculations of \citet{1996A&A...307..271S} for the interstellar cloud models with $A_V=1$---$5^m$ demonstrate that C+ is present mainly in the medium between clumps while the clumps contain CO molecules. Also optical properties of dusty clump significantly affect the clump thermal balance and its temperature \citep[see e.g.][]{Cubick_ea08}. This should influence the rates of gas and grain chemical reactions and desorption processes. Second, it is important to constrain the properties of the proposed compact dusty clumps that do not contradict the well--known average optical properties of diffuse ISM (interstellar extinction on the line of sight in the first place).

A number of researches worked on the problem of radiative transfer in inhomogeneous clumpy medium. The numerical models include those that utilize the Monte-Carlo methods \citep[see e.g.][]{1996ApJ...463..681W, 1998A&A...340..103W, 2000ApJ...528..799W} or treat the radiative transfer in the clumpy medium as a stochastic process \citep[e.g.][]{1990A&A...228..483B, 2003MNRAS.342..453H}. The analytical approximations that take into account the scattering of radiation by clumps treat clumps as large grains or mega-grains \citep[e.g.][]{1991ApJ...370L..85N, 1993MNRAS.264..145H, 1993MNRAS.264..161H, 1999ApJ...523..265V}. A simpler model is introduced by \citet{1984ApJ...287..228N}, which provides simple analytic approximations for the effective optical depth of the clumpy medium without taking into account the scattering and assuming an empty medium between clumps. \citet{1999ApJ...523..265V} show that the results that can be obtained with mega-grains and \citet{1984ApJ...287..228N} approximations are in a good agreement. Moreover, \citet{2015A&A...584A.108S}demonstrate that for relatively small effective optical depths, which are characteristic of diffuse ISM medium, the effect of clumps on the extinction curve due to scattering can be neglected. The model of \citet{1984ApJ...287..228N} has been discussed by~\citet{Nenkova_2002, 2008ApJ...685..147N} and extended by \citet{Conway_2018} whose formalism takes into account different possible distributions of the clump properties.

In this study, we use the formalism of \citet{Conway_2018} to explore the impact of compact dusty clumps proposed in~\citet[][]{Tsytovich_ea14} on the optical extinction of diffuse ISM as it will be observed within the diffuse region and by a distant observer. \citet[][]{Tsytovich_ea14} gives an upper limit on the size of dusty clumps. Thus in this paper, we explore the dependence of optical extinction and physical conditions inside the clumps on the clump size. Since diffuse regions can be of different geometries, and due to the fact that geometry is important for radiative transfer, we explore several cases: spherical region, cylindrical region which may be considered as an approximation to the filamentary shape, and an infinite plane-parallel slab of finite thickness. Also we perform calculations of thermal balance of a spherical dusty clump in the clumpy model of diffuse region, which is important for studies of chemical evolution in clumpy medium.

\section{Model description}

\begin{figure}
\centering
\includegraphics[width=0.99\linewidth]{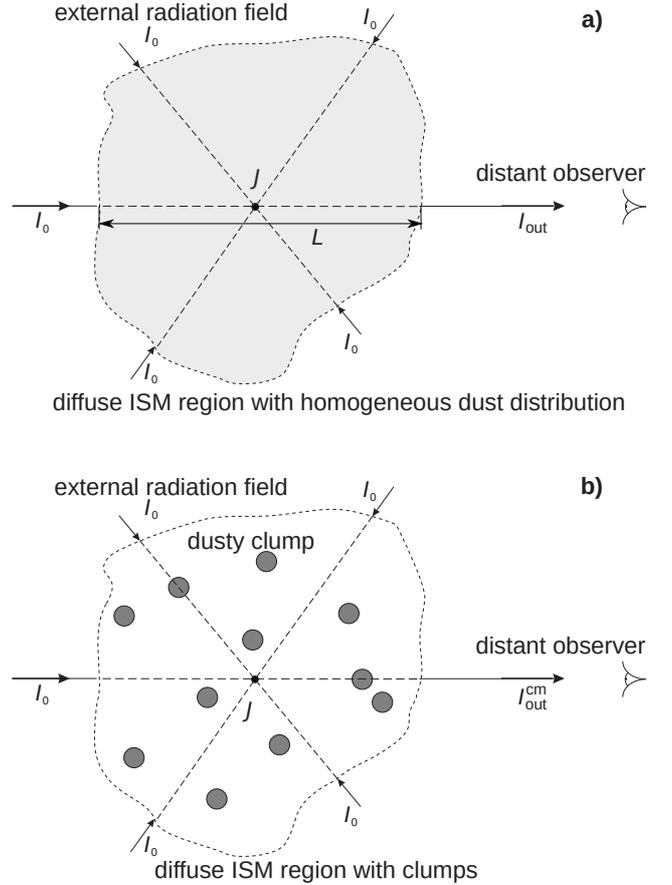}

\caption{The general scheme of the diffuse ISM region models considered in this study. The region is illuminated by isotropic external radiation field with intensity $I_0$. The black dot within the region denotes an observer within the ISM region which detects radiation with the mean intensity $J$ passed through the region. The distant observer detects the radiation passed through the region along the path with length $L$ with intensity $I_{\rm{out}}$ in the  homogeneous model (panel a) and $I_{\rm{out}}^{\rm{cm}}$ in the clumpy model (panel b). Grey colour denotes the dust grains within the ISM region in the homogeneous and clumpy models.}\label{model:fig1_1}
\end{figure}

We consider a region of the diffuse ISM isotropically illuminated by external starlight with intensity $I_{0}$ integrated over $V$-band and mean intensity $4\pi J_{0}=\int_{\Omega}I_{0}\,d\Omega = 4\pi I_0$, where $\Omega$~--- solid angle from which external starlight illuminates the region and that is equal to $4\pi$ in our case (see Figure~\ref{model:fig1_1}). There are no internal radiation sources in the region. As a reference, we consider the model with homogeneous dust spatial distribution. We investigate how the region parameters such as an extinction of external radiation field and dust temperature, which are important for astrochemical modelling, will change if all dust grains in the reference model will be grouped in spherical clumps. We consider only spherical clumps embedded in the diffuse ISM region (denoted by dotted line in Figure~\ref{model:fig1_1}b) that can have arbitrary geometries. Hereafter, the model of ISM in the region with homogeneous dust distribution will be referred to the homogeneous model, while the model of ISM region with clumps will be referred to the clumpy model.

The following are the main input parameters for these models.
\begin{itemize}
  \item $A_V^{\rm{obs}} = -1.086\ln\left(I_{\rm{out}}/I_0 \right)$~--- starlight extinction in $V$-band for the homogeneous model for a distant observer which detects the radiation with intensity $I_{\rm{out}}$. This parameter characterizes the medium optical properties along a specific line of sight only. No assumptions on the geometry of diffuse region are invoked here. In this study we focus on the diffuse molecular medium. The diffuse molecular ISM is characterized by $A_V^{\rm{obs}}\simeq 0.3$~---~$1^{m}$~\citep{2006ARAA..44..367S}. In this study, we consider $A_{V}^{\rm{obs}}=0.3$ and $1.086^{m}$. Note, that we consider $A_V^{\rm{obs}}$ as some typical value for the region obtained along a line of sight. The orientation and position of this line of sight with respect to the region borders depend on the region geometry and distant observer position. We will consider different geometries in Section~\ref{opacity_of_region}.
  \item $V_{\rm{c}}/V$~--- clump volume filling factor that is the ratio of the total volume of dusty clumps in the region to the entire region volume. This factor can be estimated taking into account equality of the total dust mass in the homogeneous and clumpy models
\begin{align}
  m_{\rm{d}}\,n_{\rm{d}}\,V&=m_{\rm{d}}\,n_{\rm{d}}^{\rm{c}}\,V_{\rm{c}}\,, \label{model:mass_conserv}
\end{align}
\begin{align}
  \frac{V}{V_{\rm{c}}}&=\frac{n_{\rm{d}}^{\rm{c}}}{n_{\rm{d}}}\,, \label{model:n_ratio}
\end{align}
where $m_{\rm{d}}$~--- mass of a dust grain particle; $n_{\rm{d}}$ and $n_{\rm{d}}^{\rm{c}}$~--- dust particles number density for the homogeneous and clumpy model (dust particles number density in clumps only), respectively. \citet{Tsytovich_ea14} estimate maximum $n_{\rm{d}}^{\rm{c}}/n_{\rm{d}}\simeq10^7$ at the clump centre. The dust number density radial distribution within a clump depends significantly on the dust charge, ion, and electron densities at the clump centre which are the free parameters of \citet{Tsytovich_ea14} model. For simplicity, we assume that the dust number density within a clump is constant and $n_{\rm{d}}^{\rm{c}}/n_{\rm{d}}=10^6$, which corresponds to the average dust number density in a clump. We, therefore, will consider the models with $V_{\rm{c}}/V=10^{-6}$. In Appendix~\ref{sec:nonuni}, we estimate that the optical properties of clumpy medium do not change significantly if one consider non-uniform dust density distribution within clumps. The spatial distribution of clumps in the clumpy model is assumed to be uniform.
  \item $n_{p}=n_{\rm{H}}+n_{\rm{H2}}=100$~cm$^{-3}$~--- total hydrogen number density in the region~\citep{2006ARAA..44..367S,2013ApJ...774..134G}.
  \item $R_{\rm{c}}$~--- typical radius of dusty clumps in the clumpy model. According to~\citet{Tsytovich_ea14}, $R_{\rm{c}}$ depends on physical conditions and dust grain properties in a diffuse ISM region. In particular, the upper limit for clump radius is inversely proportional to the gas number density in ISM. From the estimates given by \citet{Tsytovich_ea14} for the reflection nebula and warm neutral medium ISM phases, it follows that the upper limit of the clump radius is $\sim$0.1 au for $n_{p}=10^2$ cm$^{-3}$ and negatively charged grains. In the diffuse molecular medium, the external radiation is intense and hard enough so that the grains will be positively charged. Considering that the upper limit of clump radius for positively charged grains is several times larger than for negatively charged grains \citep{Tsytovich_ea14} and the uncertainties of \citet{Tsytovich_ea14} model parameters, we set this limit to 1~au for the diffuse ISM conditions considered in this study. Taking into account possible variations of physical conditions and the size distribution of dust grains in a~diffuse ISM region, one can expect that the clump radii distribution within a~region will be non-uniform. However, the exact shape of this distribution is unknown. We therefore assume that all clumps within the ISM region have the same size and consider $R_{\rm{c}} \leq1$~au.
\end{itemize}
Using the parameters above and the ratio between the total hydrogen column density, N$_{\rm{H}}$, and $A_V^{\rm{obs}}$ of N$_{\rm{H}}/A_V^{\rm{obs}}=1.8\times$10$^{21}$~cm$^{-2}$mag$^{-1}$ from \citet{1978ApJ...224..132B} \footnote{obtained with typical $R_V=3.1$ \citep{1975A&A....43..133S}}, one can estimate the typical size of the diffuse ISM region along a line of sight for a distant observer who detects $I_{\rm{out}}$ in the homogeneous model, $L$, as
\begin{equation}
L=\frac{\rm{N}_{\rm{H}}}{n_p}=1.8\times 10^{21}\,\frac{A_V^{\rm{obs}}}{n_{p}}. \label{model:eqL}
\end{equation}
The values of $L$ are $1.75$ and $6.33$ pc for $A_{V}^{\rm{obs}}=0.3$ and $1.086^{m}$, respectively.

To assess the optical depth of a single clump and of the whole ISM region in the clumpy model in the following sections we assume that the region size and the dust mass within the region with clumps are the same as in the homogeneous model.

\section{Opacity of a single dusty clump}
\label{sec:clump_opac}

In the clumpy model, chemical reactions on the dust grains are determined by the radiation field density and dust temperature within the dusty clumps which depend on the opacity of clumps for radiation. In this section, we estimate the optical depth for extinction in the centre of a dusty clump in $V$-band, $\tau_{\rm{c}}$, for clumps of different radii in the clumpy model. We start from the following expressions~\citep[see e.~g.][]{2010pcim.book.....T}
\begin{align}
  \tau_{\rm{d}}&=\frac{A_V^{\rm{obs}}}{1.086}=\int\limits_{0}^{L}n_{\rm{d}}\,\sigma_{\rm{d}}\,dr=n_{\rm{d}}\,\sigma_{\rm{d}}\,L\,, \label{model:taud}\\
  \tau_{\rm{c}}&=\int\limits_{0}^{R_{\rm{c}}}n_{\rm{d}}^{\rm{c}}\,\sigma_{\rm{d}}\,dr= n_{\rm{d}}^{\rm{c}}\,\sigma_{\rm{d}}\,R_{\rm{c}}\,, \label{model:tauc}
\end{align}
where
$\tau_{\rm{d}}$~--- dust optical depth for extinction of external starlight in $V$-band for the homogeneous model (see Figure~\ref{model:fig1_1} panel (a)); $\sigma_{\rm{d}}$~--- effective dust extinction cross-section; $n_{\rm{d}}$ and $n_{\rm{d}}^{\rm{c}}$~--- dust particles number density for the homogeneous and clumpy model (dust particles number density in clumps only), respectively. Note, that we use $R_{\rm{c}}$ in the integral upper limit in equation~\eqref{model:tauc} instead of $2R_{\rm{c}}$ because we need the optical depth to the centre of a clump.

Using equations \eqref{model:taud} and \eqref{model:tauc}, the following ratio can be derived
\begin{align}
  \frac{\tau_{\rm{c}}}{\tau_{\rm{d}}}=\frac{n_{\rm{d}}^{\rm{c}}}{n_{\rm{d}}}\,\frac{R_{\rm{c}}}{L}\,. \label{model:tauc_rat}
\end{align}

Combining equations ~\eqref{model:n_ratio}, \eqref{model:eqL}, and \eqref{model:tauc_rat} we get the following final expression
\begin{align}
  \tau_{\rm{c}}=\tau_{\rm{d}}\,\frac{V}{V_{\rm{c}}}\,\frac{R_{\rm{c}}}{L} = 6.03\times 10^{-22} R_{\rm{c}}\,n_{p} \frac{V}{V_{\rm{c}}}\,, \label{model:tauc_rat_fin}
\end{align}
which can be used to calculate $\tau_{\rm{c}}$ for given values of $n_p$, $R_{\rm{c}}$ and $V_{\rm{c}}/V$.

Note that $\tau_{\rm{c}}$ does not depend on the geometry of the diffuse ISM region in the clumpy model and $A_V^{\rm{obs}}$. From Table~\ref{model:tab1}, it is seen that $\tau_{\rm{c}}$ is close to unity for the clumps with radii of about 1 au. Absorption efficiency~\citep[see e.~g.][]{2010pcim.book.....T} depends on the radiation wavelength, $\lambda$, as $\lambda^{-x}$ where $1<x<2$. Together with our estimates of $\tau_{\rm{c}}$ in $V$-band, this implies that a single clump with $R_{\rm{c}} > 0.1$~au can be opaque for external ultraviolet radiation but still transparent for self infrared radiation.

\begin{table}
\caption{Optical properties of a single dusty clump depending on its radius. $2A_{V}^{\mathrm{c}}=2\times 1.086\,\tau_{\mathrm{c}}$~--- maximum value of starlight extinction in V-band of a clump for a distant observer.}\label{model:tab1}
  \centering
  \begin{tabular}{rrr}
  \hline
  $R_{\mathrm{c}}$, au & $\tau_{\mathrm{c}}$ & $2A_{V}^{\mathrm{c}}$, mag \\
  \hline
 0.1 & 0.09 & 0.2\\
 0.5 & 0.45 & 0.98\\
   1 & 0.9 & 1.96 \\
  \hline
  \end{tabular}

\end{table}

\section{Opacity of diffuse ISM region in the homogeneous and clumpy models}\label{opacity_of_region}

Modelling of chemical processes in the interstellar medium requires the value of radiation field energy density $u$ for photochemical reactions:
\begin{align}
u=\frac{4\pi}{c}\,J \,, \label{model:raddensity}
\end{align}
where $J$ is the mean intensity and $c$ is the speed of light. In practice, rate coefficients for photo-processes are parametrized through the extinction in $V$-band \citep[see e.g.][]{2012ApJS..199...21W}. Thus, in order to model the chemical composition of a diffuse ISM region it is necessary to estimate extinction of external starlight within this region.

We estimate the extinction of external starlight in $V$-band, $A_{\rm{V}}$, as it will be seen by the observer located within the ISM region (see also~\citet{1984ApJ...287..228N})
\begin{align}
A_{\rm{V}} = -1.086\,\ln\left(\frac{4\pi J}{4\pi J_{0}}\right)
 =  -1.086\,\ln\left(\frac{1}{4\pi I_{0}}\int\limits_{\Omega} I(\Omega) \,d\Omega\right) \,, \label{model:Av_cm_def}
\end{align}
where $I(\Omega)$ and $J$ are, respectively, intensity and mean intensity of incident radiation integrated over $V$-band seen by the observer within the ISM region. The extinction of this kind can not be measured directly by a distant observer. Note that in the clumpy model case, we assume that the observer in the ISM region is not within a clump. In the case if the observer is within a clump, the extinction is a sum of extinctions by the clump $1.086\tau_{\rm{c}}$ and extinction of the clumpy medium obtained in this Section (see below). We stress that, as in the case of astrochemical models, we consider photons from all directions but not from some given direction. In the spherical coordinate system ($r$, $\theta$, $\phi$) centred at the observer within the ISM region, equation~\eqref{model:Av_cm_def} becomes
\begin{align}
A_{\rm{V}} = -1.086\,\ln\left(\frac{1}{4\pi I_0}\int\limits_{0}^{2\pi}\int\limits_{0}^{\pi} I(\theta,\phi) \sin\theta \,d\theta\,d\phi\right) \,. \label{model:Av_cm_sph}
\end{align}

In the case of the homogeneous model, assuming that the propagating radiation does not affect the dust extinction, the intensity of radiation through the medium can be calculated as

\begin{align}
I(\theta,\phi) = I_{0} \exp\left( - \tau(\theta,\phi) \right) \,, \label{model:I_def_hom}
\end{align}
where $\tau(\theta,\phi)$ is an optical depth for extinction for a given line of sight. The optical depth $\tau(\theta,\phi)$ can be estimated from the following ratio

\begin{align}
\frac{\tau(\theta,\phi)}{s(\theta,\phi)} = \frac{\tau_{\rm{d}}}{L} = \frac{A_{V}^{\rm{obs}}}{1.086\,L} \,, \label{model:tau_hom}
\end{align}
where $s(\theta,\phi)$ is the distance between the observer within the ISM region and the region border along a given line of sight. Taking into account equations~\eqref{model:Av_cm_sph}, \eqref{model:I_def_hom}, and~\eqref{model:tau_hom}, the extinction of the external radiation field in the homogeneous model, $A_{\rm{V}}^{\rm{h}}$, can be expressed as

\begin{align}
A_{\rm{V}}^{\rm{h}} = -1.086\,\ln\left(\frac{1}{4\pi}\int\limits_{0}^{2\pi}\int\limits_{0}^{\pi} \exp\left[ -\frac{A_{V}^{\rm{obs}}}{1.086\,L}\,s(\theta,\phi) \right] \sin\theta \,d\theta\,d\phi\right) \,. \label{model:Av_hom_gen}
\end{align}

Following \citet{1984ApJ...287..228N}, the intensity of radiation through the clumpy medium can be calculated as
\begin{align}
I(\theta,\phi) = I_{0} \exp\left( - \tilde{\tau}(\theta,\phi) \right) \,, \label{model:I_def}
\end{align}
where $\tilde{\tau}(\theta,\phi)$ is an effective optical depth for extinction corresponding to the optical depth of a uniform layer which would cause the same extinction as the actual ISM region with clumps on a given line of sight. As it was noted by~\citet{2008ApJ...685..147N}, equation~\eqref{model:I_def} is valid only in a statistical sense as the intensity $I(\theta,\phi)$ is an average over an ensemble of all possible clump positions along the same line of sight. \citet{1984ApJ...287..228N} suggested that in the case of random clump positions one can use Poisson distribution to estimate $\exp\left(\tilde{\tau}(\theta,\phi)\right)$. \citet{Conway_2018} generalized the formulation from \citet{1984ApJ...287..228N} to the case of a medium comprised of a mixture of clumps with different properties and taking into account the clump geometry. With Monte-Carlo simulations, \citet{Conway_2018} show that their formalism provides the estimates of clumpy medium absorption with the accuracy of better than 15 per cent for $V_{c}/V \lesssim 0.1$. We consider significantly lower value of $V_{c}/V=10^{-6}$ and, therefore, the formalism of \citet{Conway_2018} should provide us the accurate estimates of the ISM region extinction. Below, we estimate the accuracy of this formalism with Monte-Carlo calculations for the region parameters that have not been considered in Monte-Carlo calculations of \citet{Conway_2018} (see Section~\ref{method_sph_sphere}). Employing equation~(23) of~\citet{Conway_2018}, equation~\eqref{model:I_def} can be transformed into

\begin{align}
I(\theta,\phi) = I_{0} \exp\left[ - \mathcal{N}(\theta,\phi) \left( 1- \left<e^{-2\tau_{\rm{c}}} \right> \right) \right] \,, \label{model:I_def_clumpy}
\end{align}
where $\mathcal{N}(\theta,\phi)$ is the average number of clumps along a given line of sight; $\left<e^{-2\tau_{\rm{c}}} \right>$ is the average of $e^{-2\tau_{\rm{c}}}$ over a clump projected area in the case of spherical clumps or over a clump projected area and all clump orientations for non-spherical clumps. The averaging in $\left< e^{-2\tau_{\rm{c}}} \right>$ should be performed for rays that originate in the observer's position within the ISM region, pass through a clump and are not exactly parallel. However, one can assume that these rays are parallel taking into account that the clump radius is significantly smaller than the region size and, thus, most of the clumps have a small angular size. For parallel rays, we can use equation~(35b) from~\citet{Conway_2018} \citep[see also][]{Ignace2004} and approximate $\left< e^{-2\tau_{\rm{c}}} \right>$ as
\begin{align}
\left< e^{-2\tau_{\rm{c}}} \right> = \frac{2}{\left(2\tau_{\rm{c}}\right)^2}\left[ 1-(1+2\tau_{\rm{c}})\exp(-2\tau_{\rm{c}}) \right] \,. \label{model:mean_exp}
\end{align}
\citet{Conway_2018} show that $\tilde{\tau}(\theta,\phi)$ and $\tau(\theta,\phi)$ are related through the clumping correction factor
\begin{align}\label{correct_factor}
K(\tau_{\rm{c}}) &= \frac{\tilde{\tau}(\theta,\phi)}{\tau(\theta,\phi)}= \frac{1- \left<e^{-2\tau_{\rm{c}}} \right>}{\left<2\tau_{\rm{c}} \right>} = \nonumber\\
&=\frac{3}{2}\,\frac{1}{2\tau_{\rm{c}}}\,\left(1-\frac{2}{\left(2\tau_{\rm{c}}\right)^2}\left( 1-\left(1+2\tau_{\rm{c}}\right)\exp\left(-2\tau_{\rm{c}}\right) \right)\right)\,,
\end{align}
where $\left<2\tau_{\rm{c}} \right>$ is $2\tau_{\rm{c}}$ averaged in the same way as $\left<e^{-2\tau_{\rm{c}}} \right>$ and can be approximated with equation~(35a) from~\citet{Conway_2018}, assuming that the rays passing through clumps are parallel. The correction factor $K(\tau_{\rm{c}})$ is always $\leq 1$ and does not depend on the region geometry and size.

The average number of clumps on a given line of sight can be estimated as \citep[see e.g.][]{Conway_2018}
\begin{align}
\mathcal{N}(\theta,\phi)= \int\limits_{0}^{s(\theta,\phi)} n_{\rm{c}} \, A \, dr = \int\limits_{0}^{s(\theta,\phi)} n_{\rm{c}} \, \pi R_{\rm{c}}^2 \, dr \,, \label{model:Nave_def}
\end{align}
where $n_{\rm{c}}$~--- number density of clumps in the ISM region; $A$~---clump area perpendicular to a line of sight.

The clumps number density and total number of clumps in the diffuse ISM region, $N_{\rm{c}}$, can be found using the following relation based on \eqref{model:mass_conserv}
\begin{align}
m_{\rm{d}}\,n_{\rm{d}}\,V&=m_{\rm{d}}\,n_{\rm{d}}^{\rm{c}}\,N_{\rm{c}}\frac{4}{3}\,\pi R_{\rm{c}}^3\,. \label{model:mass_cons_Nc}
\end{align}
From equations \eqref{model:n_ratio} and \eqref{model:mass_cons_Nc}, we find clumps number density
\begin{align}
n_{\rm{c}}=\frac{N_{\rm{c}}}{V}& = \frac{3}{4\pi R_{\rm{c}}^3} \frac{V_{\rm{c}}}{V}\,. \label{model:Nc_common}
\end{align}
With \eqref{model:Nc_common} and taking into account the uniform clump spatial distribution, equation \eqref{model:Nave_def} becomes
\begin{align}
\mathcal{N}(\theta,\phi) = \frac{3}{4R_{\rm{c}}} \frac{V_{\rm{c}}}{V} \, \int\limits_{0}^{s(\theta,\phi)} \, dr = \frac{3}{4R_{\rm{c}}} \frac{V_{\rm{c}}}{V} \, {s(\theta,\phi)}\,. \label{model:Nave}
\end{align}

Using \eqref{model:Av_cm_sph}, \eqref{model:I_def_clumpy}, and \eqref{model:Nave} we find the following final expression for extinction within the diffuse ISM region with uniformly distributed spherical clumps
\begin{align}
A_{\rm{V}}^{\rm{cm}} = -1.086\,\times \nonumber\\
\ln \left(\frac{1}{4\pi }\int\limits^{2\pi }_{0}\int\limits^{\pi }_{0} \exp\left[ -\frac{3}{4R_{\rm{c}}} \frac{V_{\rm{c}}}{V} \, \left( 1 - \left<e^{-2\tau_{\rm{c}}} \right> \right) \, s(\theta,\phi) \right] \sin\theta\, d\theta\, d\phi \right)\,,\label{model:Av_clumpyreg}
\end{align}
where $\tau_{\rm{c}}$ and $\left<e^{-2\tau_{\rm{c}}} \right>$ are given by equations \eqref{model:tauc_rat_fin} and \eqref{model:mean_exp}, respectively. The equations \eqref{model:Av_hom_gen} and \eqref{model:Av_clumpyreg} can be applied for an arbitrary geometry of the ISM region and the observer position within the region, which are determined by $s(\theta,\phi)$. The geometry of diffuse ISM regions is unknown. In the general case of arbitrary ISM region shape, the calculations of $A_{\rm{V}}^{\rm{h}}$ and $A_{\rm{V}}^{\rm{cm}}$ can be carried out only numerically. However, for simple geometries, analytical or semi-analytical estimations are possible. Below, we derive expressions for $A_{\rm{V}}^{\rm{h}}$ and $A_{\rm{V}}^{\rm{cm}}$ in the case of spherical, slab and filamentary geometries of the diffuse ISM region.

\subsection{Sphere with radius $L/2$.}\label{method_sph_sphere}

Spherical region is the simplest approximation for a cloud-like object. We consider the spherical ISM region with radius of $L/2$.

For the observer located in the region centre, the analytical result for $A_{V}^{\rm{h}}$ and $A_{V}^{\mathrm{cm}}$ can be readily obtained as the distance from the observer to the region border does not depend on the direction and $s(\theta,\phi) = L/2$ yielding $A_{V}^{\rm{h}}=A_{V}^{\rm{obs}}/2$ and

\begin{align}
A_{\rm{V}}^{\rm{cm}} = 1.086 \times \frac{3L}{8R_{\rm{c}}} \frac{V_{\rm{c}}}{V} \left(1 - \left<e^{-2\tau_{\rm{c}}} \right> \right) \,.\label{model:Av_an_sph}
\end{align}
Therefore, the extinction in the homogeneous model is $A_{V}^{\rm{h}}=0.15$ and $0.543^m$ for $A_{V}^{\rm{obs}}=0.3$ and $1.086^m$, respectively. From Figure~\ref{model:op:fig1} and Table~\ref{model:tabavcmsym} (see Appendix), it is seen that the extinction within the ISM region decreases when all dust grains are grouped in clumps. The decrease reaches a factor of 1.8 for $R_{\rm{c}}=1$~au in comparison to the homogeneous model. The effect of clumps on the extinction within the region is significantly less pronounced in the case of smaller clumps. The difference between extinction in the clumpy and homogeneous model is less than 5 per cent for clumps with $R_{\rm{c}}<0.1$~au. The optical depth of the largest clumps is $\sim1$ (see Section~\ref{sec:clump_opac}), and, therefore, $A_{\rm{V}}^{\rm{cm}}$ is determined mainly by the average number of clumps and is always lower than the extinction of the homogeneous medium~\citep[see][]{Conway_2018}. Thus, low extinction in our clumpy model for the largest clumps is related to low average number of clumps on a line of sight $\mathcal{N}$ which, for example, is of 0.49 for $R_{\rm{c}}=1$~au and $A_V^{\rm{obs}}=1.086^m$. The value of $A_{\rm{V}}^{\rm{cm}}$ increases with decreasing clump radius because the average number of clumps on a line sight in our model is inversely proportional to $R_{\rm{c}}$. As clumps become optically thin and their number on a line of sight increases with decreasing $R_{\rm{c}}$, the medium absorption properties approach the smooth-density limit in accordance to the results of \citet{Conway_2018}.

In order to check our analytical estimates of $A_V^{\rm{cm}}$, we carried out Monte-Carlo calculations of the opacity of clumpy medium. We considered the rays directed from the region centre. The rays direction was random and the number of rays, $N_{\rm{rays}}$, was of the order of $10^4$---$10^5$. The initial intensity of radiation along a ray was assumed to be one arbitrary unit. In the case, if the ray intersects a clump, the radiation intensity along this ray became $I_i = I_i \exp(-2\tau_{\rm{c}}d/(2R_{\rm{c}}))$, where $d$ is the distance that the ray passed through the clump. The spatial distribution of clumps was random and uniform within the sphere with radius $L/2$. The number of clumps in the clumpy model is large ($>3\times10^{11}$) that makes Monte-Carlo calculations computationally intensive. For the sake of calculations speed, the clump positions were generated with the default 64 bit variant of random number generator from the PCG family\footnote{\url{http://www.pcg-random.org}}.
The resulting value of $A_V^{\rm{cm}}$ was estimated as
\begin{equation}
A_V^{\rm{cm}} = -1.086 \ln \left( \frac{\sum\limits_{i=1}^{N_{\rm{ph}}} I_i }{N_{\rm{rays}}} \right)\,,
\end{equation}
where summation goes over the rays. The results of Monte-Carlo simulations are shown in Figure~\ref{model:op:fig1}. The difference between Monte Carlo analytical estimates is no more than 4 percent for the largest clump radius considered in this study. Such a difference is significantly lower than the uncertainty in the $A_V^{\rm{obs}}$ values for diffuse ISM or in the clump parameters. Therefore, we conclude that \citet{Conway_2018} formalism allows us to obtain a good estimate of the clumpy medium optical properties.

\begin{figure}
\centering
\includegraphics[width=1.0\linewidth]{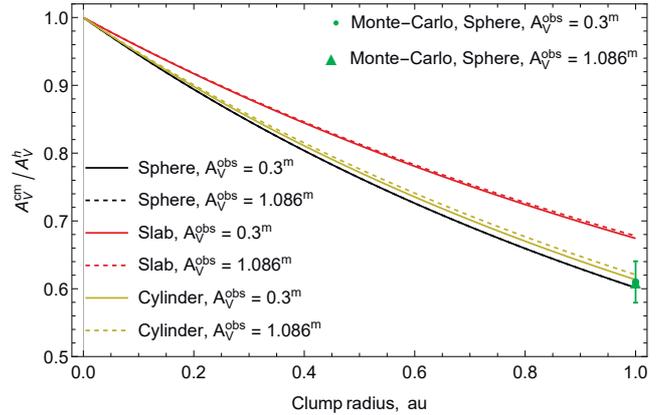}
\caption{The ratio of extinction in the clumpy and homogeneous ISM region models as a function of clump size, for a given $A_{V}^{\rm{obs}}$ and different region geometries: in the centre of spherical region (black), in the midplane of infinite slab (red), and on the axis of infinite cylinder (yellow). Analytical estimates are shown along with the Monte-Carlo simulation results for the spherical region only (green). Dispersion of Monte-Carlo simulations does not exceed 5 per cent.}\label{model:op:fig1}
\end{figure}

\subsection{Infinite slab with thickness $L$}

\begin{figure}
\centering
\includegraphics[width=0.7\linewidth]{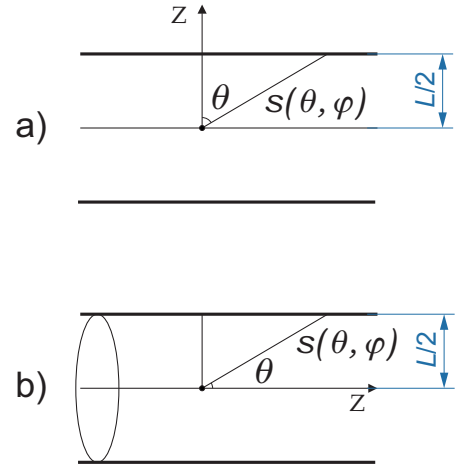}
\caption{Geometry of diffuse region with clumps. (a) Plane parallel slab with thickness $L$. (b) Infinite cylinder with radius $L/2$. The black dot denotes the observer within the diffuse region.}\label{model:figclregions}
\end{figure}

In this section, we estimate the extinction calculated for the observer in the middle plane of an infinite plane parallel slab with thickness $L$ (see Figure~\ref{model:figclregions}, panel~a)). In this case, $s(\theta,\phi) = L/\left(2\cos\theta\right)$. Replacing $x=1/\cos \theta$, the extinction of the external radiation field in the middle plane is
\begin{align}
A_V^{\rm{h}} = -1.086\,\ln\left(\int\limits_{1}^{\infty} \exp\left[ -\frac{A_{V}^{\rm{obs}}}{2.172} \,x \right]\,\frac{dx}{x^{2}} \right) = \nonumber\\
 = -1.086\,\ln\left[ \exp(-k)-k\,E_1(k)  \right] \,  \nonumber\\
k \equiv \frac{A_{V}^{\rm{obs}}}{2.172} \,,\\
A_V^{\rm{cm}} = -1.086\,\ln\left(\int\limits_{1}^{\infty} \exp\left[ -\frac{3L}{8R_{\rm{c}}} \frac{V_{\rm{c}}}{V} \left(1 - \left<e^{-2\tau_{\rm{c}}} \right> \right) \,x \right]\,\frac{dx}{x^{2}} \right) = \nonumber\\
 = -1.086\,\ln\left[ \exp(-t)-t\,E_1(t)  \right] \nonumber\\
t \equiv \frac{3L}{8R_{\rm{c}}} \frac{V_{\rm{c}}}{V} \left(1 - \left<e^{-2\tau_{\rm{c}}} \right> \right) \,, \label{model:Av_0_approx_avcm_slab}
\end{align}
where $E_1$ is the first exponential integral. The values of $A_V^{\rm{h}}$ for $A_{V}^{\rm{obs}}=0.3$ and $1.086^m$ are 0.45 and $1.21^m$, respectively.

As in the case of a spherical ISM region, the extinction in clumpy model is lower than in the homogeneous model (see Figure~\ref{model:op:fig1}). The decrease of extinction due to clumps is only slightly lower than in the case of spherical region and it approaches 30 per cent for $R_{\rm{c}}=1$~au, in comparison to the homogeneous model. At the same time, slab geometry shields external radiation field most effectively among considered geometries in a sense that the absolute values of extinction are higher for all clump radii among considered geometries (see Table~\ref{model:tabavcmsym}).

\subsection{Infinite cylinder with radius $L/2$}

An infinite cylinder can be considered as representative of filamentary diffuse ISM region. We calculate the extinction for the observer located on the axis of infinite cylinder with radius $L/2$ (see Figure~\ref{model:figclregions}, panel~b)). In this case, $s(\theta,\phi) = L/(2\sin\theta$). Replacing $x=1/\sin \theta$, extinction of the external radiation field on the cylinder axis is

\begin{align}
A_V^{\rm{h}} = -1.086\,\ln\left(\int\limits_{1}^{\infty} \exp\left[ -\frac{A_{V}^{\rm{obs}}}{2.172} \,x \right]\,\frac{dx}{x^{2}\sqrt{x^2-1}} \right) \,, \label{model:Av_0_approx_avcm_cyl1}\\
A_V^{\rm{cm}} = -1.086\,\ln\left(\int\limits_{1}^{\infty} \exp\left[ -\frac{3L}{8R_{\rm{c}}} \frac{V_{\rm{c}}}{V} \left(1 - \left<e^{-2\tau_{\rm{c}}} \right> \right) \,x \right]\,\frac{dx}{x^{2}\sqrt{x^2-1}} \right) \,. \label{model:Av_0_approx_avcm_cyl2}
\end{align}
Integration in~\eqref{model:Av_0_approx_avcm_cyl1} and \eqref{model:Av_0_approx_avcm_cyl2} can be performed numerically. For the homogeneous model, one get $A_V^{\rm{h}}=0.22$ and $0.74^m$ for $A_{V}^{\rm{obs}}=0.3$ and $1.086^m$, respectively. The difference between $A_V^{\rm{h}}$ and $A_V^{\rm{cm}}$ is almost the same as in the case of spherical ISM region (see Figure~\ref{model:op:fig1} and Table~\ref{model:tabavcmsym}). In the case of filamentary region model, $A_V^{\rm{cm}}$ is lower than $A_V^{\rm{h}}$ by $\sim40$ per cent for $R_{\rm{c}}=1$~au.

\subsection{Extinction off the region symmetry centre}

\begin{figure}
\centering
\includegraphics[width=1.00\linewidth]{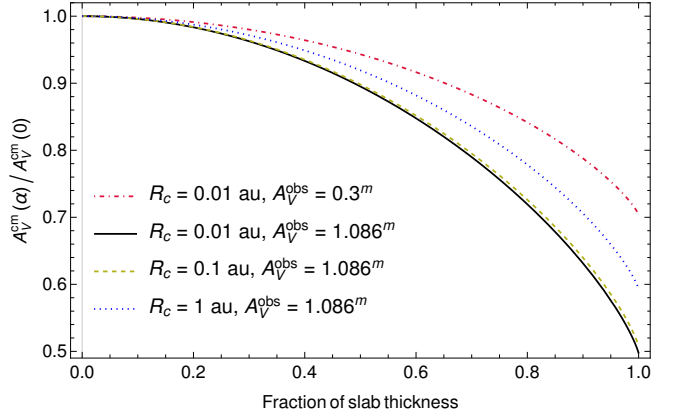}
\caption{The ratio of extinction of external stellar radiation field at different distances from the infinite slab central plane to the extinction at the slab central plane in the clumpy model for different clump radii and $A_V^{\rm{obs}}$. Fraction of slab thickness in $z$-direction $\alpha$ is equal to 1 at slab border and to 0 at slab central plane.}\label{model:op:fig2}
\end{figure}

In the previous Sections, we considered the extinction for the observer situated in the region symmetry centre, plane, and axis. In this Section, we estimate how extinction varies if the observer is not in this position. We make these estimates using the infinite slab region model. For the slab geometry, one can obtain the analytical expression for the extinction for the observer located off the slab central plane from~\eqref{model:Av_0_approx_avcm_slab}:
\begin{align}
A_V^{\rm{cm}}(\alpha) = -1.086\,( \ln[ \exp(-t(1+\alpha))-t(1+\alpha)\,E_1(t(1+\alpha)) + \nonumber\\
 \exp(-t(1-\alpha))-t(1-\alpha)\,E_1(t(1-\alpha)) ] + \ln[0.5] ) \,, \label{model:avcm_slab_vs_a}
\end{align}
where $\alpha$ is the fraction of $L/2$ defined so that the observer is located at the distance $\alpha L/2$ from the slab central plane. From this expression, we estimated how the value of $A_V^{\rm{cm}}$ changes if it is calculated off the slab central plane. We obtained that $A_V^{\rm{cm}}$ decreases by no more than 10 per cent at half way from the region central plane to its border  (see Figure~\ref{model:op:fig2}). The difference between extinction at the central and border parts of the diffuse region decreases as the region becomes more transparent due to increasing clump size and/or decreasing $A_V^{\rm{obs}}$.

We expect that the difference between $A_V^{\rm{cm}}$ in and off the ISM region symmetry centre for other region geometries considered in this study will be similar or even lower than for the slab as the slab has maximum opacity among the other geometries. Therefore, we assume that the estimates of extinction in the region symmetry centre are applicable for the chemical modelling in the most positions within the diffuse ISM region of any geometry considered in this study.

\subsection{Effective extinction for a distant observer}

We also calculated the effective extinction of the whole ISM region with clumps as it will be seen for a distant observer located outside of the ISM region ${A}_V^{\rm{eff}} = -1.086\ln\left( I^{\rm{cm}}_{\rm{out}}/I_0 \right)$, where $I^{\rm{cm}}_{\rm{out}}$ is the intensity of the emission passed through the clumpy ISM region along the path with length $L$ and detected by the distant observer (see Figure~\ref{model:fig1_1}). Note, that $I^{\rm{cm}}_{\rm{out}}$ and, thus, ${A}_V^{\rm{eff}}$ make statistical sense only being an average over many realizations of random clump positions along the same line of sight and expected to be close to an average over many line of sights within the observations beam passing through a medium with the same properties~\citep[see][]{Conway_2018}. The value of ${A}_V^{\rm{eff}}$ can be directly compared with ${A}_V^{\rm{obs}}$. With equation \eqref{model:Av_clumpyreg} and setting $s(\theta,\phi) = L\,\updelta(0,0)$, where $\updelta$ is the Dirac function, the value of ${A}_V^{\rm{eff}}$ can be estimated as

\begin{align}
A_{\rm{V}}^{\rm{eff}} = 1.086 \times \frac{3L}{4R_{\rm{c}}} \frac{V_{\rm{c}}}{V} \left(1 - \left<e^{-2\tau_{\rm{c}}} \right> \right) \,. \label{model:Av_eff}
\end{align}

Table~\ref{model:tabAvEff} shows that, as in the case of an observer within the ISM region, ${A}_V^{\rm{eff}}$ is lower than ${A}_V^{\rm{obs}}$ by $\sim40$ per cent for $R_{\rm{c}}=1$~au. The diffuse region with clumps with $R_{\rm{c}}<0.1$~au is practically indistinguishable from the diffuse region with homogeneous dust distribution for a distant observer in terms of absolute values of visible extinction.

\begin{table}
  \caption{Effective extinction of stellar background emission in $V$-band in the clumpy model as it will be observed by a distant observer.}\label{model:tabAvEff}  \centering
  \begin{tabular}{rrr}
  \hline
  $R_{\mathrm{c}}$, au & $A_V^{\rm{eff}}$ ($A_{V}^{\rm{obs}}=0.3^m)$ & $A_V^{\rm{eff}}$ ($A_{V}^{\rm{obs}}=1.086^m)$ \\
  \hline
  0.1 & 0.27 & 1.01\\
  0.5 & 0.22 & 0.81\\
  1 & 0.18 & 0.65\\
  \hline
  \end{tabular}

\end{table}

\section{Dust temperature radial distribution in a single clump}
\label{sec:td}

Our calculations show that the optical depth from the edge to the centre of clump is close to unity for clumps with $R_{\mathrm{c}}\simeq1$~au. Such clumps are opaque for external UV radiation and the dust temperature inside these clumps should be lower than dust temperature in the homogeneous medium. In this Section, we estimate the dust temperature within clumps assuming that the dust density distribution within clumps is constant. Note that in Appendix~\ref{sec:nonuni} we show, that the difference in dust temperatures in model with constant and non-uniform dust density distributions within clumps does not exceed 0.5~K.

\begin{figure}
\centering
\includegraphics[width=0.95\linewidth]{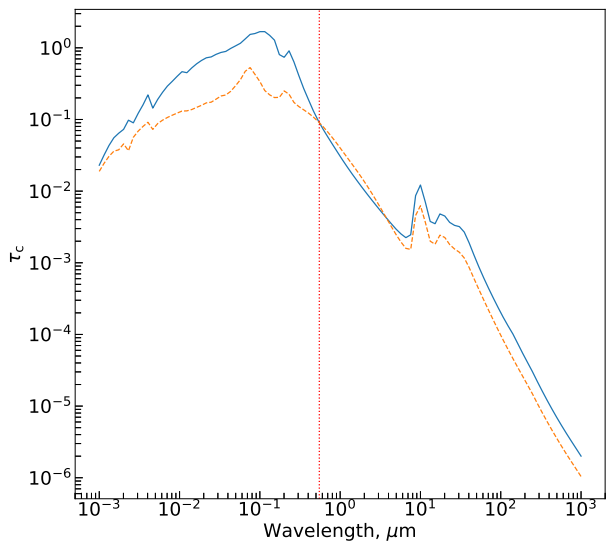}

\caption{Dependence of the clump optical depth for extinction on wavelength for $R_{\mathrm{c}}=0.1$~au. Blue solid line: optical depth calculated assuming that all dust grains have a radius of $a=0.03$~$\mu$m. Orange dashed line: optical depth calculated with dust grain sizes distribution and optical properties from~\citet{2001ApJ...548..296W}. Vertical red dotted line denotes wavelength of 5500~$\textup{\AA}$.}\label{model:dustopac:fig}
\end{figure}

We made the numerical calculations of the dust temperature distribution inside a clump with \textsc{hyperion} Monte-Carlo radiative transfer code~\citep{2011A&A...536A..79R} allowing 3D radiative transfer taking into account multiple scattering. \textsc{Hyperion} computes the dust equilibrium temperature with \citet{1999A&A...344..282L} iterative method. The clump model of \citet{Tsytovich_ea14} is developed for grains all of the same radius, $a$. The properties of dust grains, including its size, are important parameter for the dust temperature calculations. It is unknown how grain particles with different sizes will be distributed within and/or among clumps. One can speculate that the grain particles of different sizes may be contained within the same clump or that they may break into different clumps with each clump harbouring only the grains with some given size \citep[see, e.g. Section 9 in][]{Tsytovich_ea14}. Therefore, for the dust temperature calculations, we used two sets of grain properties. We assumed that all grains inside a clump have a similar size of $a=0.03$~$\mu$m, which is consistent with the grain radius values used by \citet{Tsytovich_ea14} for their estimations of clump properties. The optical properties for grains with this radius were computed with the wrapper program\footnote{\url{https://github.com/hyperion-rt/bhmie}} for the \citet{1983asls.book.....B} code using refractive indices from \citet{2003ApJ...598.1026D} and abundances from \citet{2001ApJ...548..296W}. In addition to the case of same-size grains, we also estimated the impact of optical properties of dust grains on their temperature in clumps for the case of grain size distribution taken from~\citet{2001ApJ...548..296W}\footnote{\url{http://docs.hyperion-rt.org/en/stable/dust/d03.html} dust model with $R_{V}=3.1$}. It is important to note that the dust mass density within a clump has been adjusted so that the optical depth at 5500~$\textup{\AA}$ calculated with equation \eqref{model:tauc} for a given dust extinction cross-section was equal to the optical depth calculated with equation \eqref{model:tauc_rat_fin}. The clump optical depth in the case of grains with $a=0.03$~$\mu$m is higher by a factor of $\sim4$ in the UV range ($<0.3$~$\mu$m) than the optical depth calculated with optical properties from \citet{2001ApJ...548..296W} (see Figure~\ref{model:dustopac:fig}). The calculations were performed using a spherical polar grid containing 500, 2, and 2 cells in the radial, polar and azimuthal direction, respectively. The cells widths in the radial direction were logarithmically decreased with distance from the clump centre. We set the number of photons packets to $10^8$ and the number of iterations to 100. The dust temperature variations between several last iterations did not exceed 1 per cent. The spectrum of the interstellar radiation field was taken from~\citet{1983A&A...128..212M}. The calculations were performed for both unattenuated interstellar radiation field and attenuated interstellar field with intensity decreased by a factor of $\exp(-\tau)$ where the optical depth $\tau$ has been calculated using optical properties from \citet{2001ApJ...548..296W} and normalized to be equal to ${A}_V^{\rm{cm}}/1.086$ at 5500~$\textup{\AA}$ with ${A}_V^{\rm{cm}}$ taken from Section~\ref{method_sph_sphere}. The calculations for the unattenuated interstellar radiation field are used to asses the maximum dust temperature within clumps.

In order to assess the influence of clumps, we also estimated the dust temperature in the spherical diffuse ISM region with homogeneous dust distribution and for the dust optical properties from \citet{2001ApJ...548..296W}. The model parameters were the same as in the case of calculations of dust temperature within clumps except the region size, which was calculated using the expression~\eqref{model:eqL}, and dust mass density, set so that the optical depth for extinction at 5500~$\textup{\AA}$ calculated with \eqref{model:taud} was equal to ${A}_V^{\rm{obs}}/1.086$. We obtained that the dust temperatures in the homogeneous model are in the range of 16---18~K. These results are in a good agreement with the observational estimates \citep[see e.g.][]{1996A&A...312..256B,2017ApJ...851..119R}.

\begin{figure}
\centering
\includegraphics[width=0.95\linewidth]{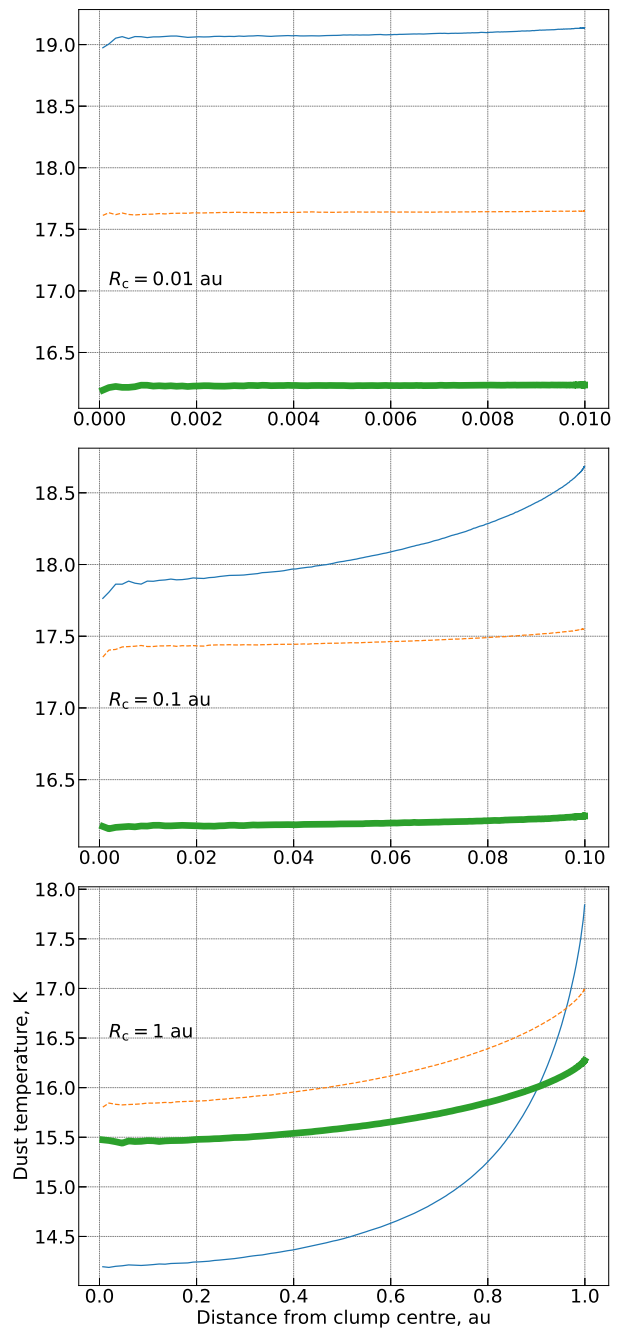}

\caption{Dust temperature profiles within clumps with different radius. Blue solid line: dust temperature calculated assuming that all dust grains have a radius of $a=0.03$~$\mu$m with unattenuated external starlight. Orange dashed line: dust temperature calculated with dust grain sizes distribution and optical properties from~\citet{2001ApJ...548..296W} with unattenuated external starlight. Green thick solid line: dust temperature calculated with dust grain sizes distribution and optical properties from~\citet{2001ApJ...548..296W} with external starlight attenuated by the clumpy medium.}\label{model:temp:fig}
\end{figure}

In figure~\ref{model:temp:fig}, it is seen that the dust temperature in clumps with $R_{\mathrm{c}}=1$~au calculated for the unattenuated interstellar radiation field decreases to the clump centre by several degrees of K and can reach 14---16~K. The dust temperature in the central regions of clumps with $R_{\mathrm{c}}=1$~au is, thus, somewhat lower than in the homogeneous diffuse region. The gradient of dust temperature decreases with decreasing clump radius and practically vanishes for clumps with $R_{\mathrm{c}}<0.1$~au. This result does not depend on whether one uses the optical properties calculated for a single dust grain radius of $a=0.03$~$\mu$m or properties from~\citet{2001ApJ...548..296W}. Note, that we do not take into account stochastic heating which can affect the dust temperature distribution within clumps. However, according to \citet[e.g.][]{2006MNRAS.367.1757C} the dust temperature variations due to stochastic heating do not depend significantly on the extinction for the grains with sizes of $\gtrsim 0.02$~$\mu$m. Therefore, for grains with sizes typically considered in astrochemical models ($\sim 0.1$)~$\mu$m, the dust temperature radial distribution in a clump does not depend significantly on the grain stochastic heating.

The dust temperatures estimated for the attenuated interstellar field are lower by no more than $\sim2$~K than those calculated for the unattenuated interstellar field (see green thick line in Figure~\ref{model:temp:fig}). The largest difference is for smallest clumps when the extinction inside a clump itself is comparable to the extinction by clumpy medium surrounding the clump.

As it was noted by \citet[equation (42)]{Tsytovich_ea14}, the ratio of the clump radius to the mean free path of atoms/molecules due to collisions with dust grains is of the order of unity. The exact value of this ratio for the diffuse ISM conditions can not be derived due to uncertainties of input parameters of \citet{Tsytovich_ea14} model. The efficiency of gas cooling strongly depends on the value of this ratio. If this ratio exceeds unity then the grain-gas coupling should be effective and the gas temperature inside a clump should be very close to the dust temperature \citep[see also][]{Ivlev_ea18}. On the other hand, if this ratio is close to unity or slightly lower then the gas cooling due to collisions with grains should be less effective and the gas temperature inside a clump can be higher than the dust temperature.

\section{Discussion and Conclusions}

We made analytical and numerical estimates of optical properties and dust temperatures in diffuse interstellar medium for the case gas---dust instability predicted by~\citet{Tsytovich_ea14}. Note that our analytical estimates of optical properties for clumpy medium can be used for the physical conditions that differ from those of diffuse interstellar medium. \citet{Tsytovich_ea14} assumed that the clumps are optically thin based on the estimates of scattering cross-section for dust grains. Our calculations show that the clumps with radii $\gtrsim0.1$~au are optically thin in the visible range but can be optically thick for far UV radiation. As it is shown by \citet{2015A&A...584A.108S}, the presence of optically thick clumps can affect the observed extinction law. The comparison of observed and model extinction laws can provide important constraints on the clump model parameters. However, in the current study, we can not predict exact effects of clumps on the extinction law as it requires substantial modifications of \citep{Tsytovich_ea14} model in order to take grain size distribution into account which should be a subject for future studies. Our calculations show that, when all dust grains in the homogeneous ISM region become grouped in large ($\sim1$~au) clumps, the region observed extinction decreases by a factor of 1.5. However the resulted clumpy region opacity still remains within the extinction limits defined by \citep[][]{2006ARAA..44..367S} for the molecular diffuse ISM. This decrease can be compensated by increasing the clumpy region length along a line of sight $L$ by a factor of 1.5 (see equation~\eqref{model:Av_eff}). The maximum clumpy region size with the effective extinction of $1^m$ along a line of sight achieves 9.5 pc for the largest clumps. To our knowledge, there are no observational data that allow us to conclude whether such a region size is feasible. The decrease can also be compensated by increasing an amount of dust in the region by a factor of 1.5 assuming that this increasing leads to a higher number of clumps in the region. Available gas-phase element abundances \citep[see e.g.][]{2004ASPC..309..393S} are lower by several orders of magnitude than required for an additional increase of the dust mass by a factor of 1.5. Therefore, the observed extinction can not be reproduced in the medium model with large clumps only by depletion of elements from the gas phase. Our results were derived from the observational estimates of the dust extinction without assumptions on the dust cross-sections for absorption and scattering. It should be noted also that our estimates do not take into account a diffuse radiation field produced due to scattering on clumps. Based on the results of \citet[][Figure 3a]{1990A&A...228..483B} we expect that the diffuse scattered radiation will not alter significantly our estimates.

The dust temperature within clumps with radii $\gtrsim0.1$~au does not exceed 20~K and can reach $\sim14$~K in central clump regions, significantly lower than the dust temperature in the diffuse ISM region with homogeneous dust distribution. Moreover, the total extinction within a clump is a sum of extinctions in the clump itself and in the clumpy medium surrounding the clump. This total extinction can be higher than the extinction within the homogeneous diffuse ISM medium by a factor of 2 reaching $1.3^m$. Such extinction and low temperatures permit formation of ice on a grain surface which can affect optical properties of the medium and chemical processes in it. For example, due to exponential dependence of thermal desorption rate on the dust temperature the difference of temperature of 3 K between the clumpy and homogeneous models will lead to difference in the thermal desorption rate for H$_2$ \footnote{H$_2$ binding energy is of 314~K} of a factor of 50 for the dust temperature of 15~K. The time needed for ice formation is of an order of a hundred thousand years \citep[see e.g.][]{2013ApJ...769...34V}. Our estimates show that the characteristic time for a dust grain particle to diffuse from clump centre to clump border exceeds several tens of thousand years for clump radii $>0.1$~au and reaches millions of years for clumps with radii $\sim1$~au (see Appendix), which is comparable to a lifetime of diffuse regions. The characteristic time for the gas particles to cross a clump is of $\sim5$ yr for 0.1 au clumps and reaches 500 years for clumps with radii $\sim1$ au (see Appendix) that is comparable to the characteristic time of gas particles accretion onto the grain surface. This means that dust grains can reside in cold regions of a clump for a time sufficient for formation of icy mantles on its surface. Together with increased gas density inside clumps \citep[see e.g.][]{Ivlev_ea18}, this makes clumps efficient `chemical reactors' that could enrich diffuse ISM with products of grain surface chemical synthesis. Molecules formed in ice mantles can be released into the gas phase via non-thermal processes such as chemical desorption or photodesorption. The molecules released from the grain surface can be destroyed then by the external stellar radiation field in the inter-clump medium which, as it follows from our calculations for different ISM region geometries, can be significantly stronger than the radiation field within the diffuse ISM region with the homogeneous dust distribution. For example, the CO photodissociation rate, which exponentially depends on the medium extinction, can be higher by 70 per cent in the gas between clumps in comparison to the homogeneous model. The stronger radiation field within the diffuse ISM region with clumps can also lead to a higher ionization degree in the medium within the region in comparison to the region with the homogeneous dust distribution. Thus, it is not immediately clear if the possible assembly of dust into small clumps in diffuse ISM should enrich or make more poor its chemical composition in comparison to the case of uniform diffuse ISM. Overall, our calculations show that the conditions in the diffuse ISM region with clumps larger than 0.1~au can affect rates of chemical processes when compared to the case of the homogeneous dust distribution. The exact effects of the presence of clumps on the diffuse region chemical composition should be studied by means of astrochemical modelling which also can be used to put additional constraints on the clump properties.

As it was shown by \citet{2014P&SS..100...32W}, the flat extinction in far mid-infrared range observed towards diffuse regions can be explained by the presence of micron-sized grains. According to study of \citet{2019NatAs.tmp..319H} such large dust grains can be destroyed by suprathermal rotation driven by radiative torques. Thus, potentially it can be problematic to explain the presence of large grains in the diffuse medium with intensive radiation field. However, the calculations of \citet{2019ApJ...876...13H} show that such grains are able to survive when the radiation intensity is lower or the local gas density is higher than the average ones for the diffuse ISM. The clumps with radii $>0.1$~au can provide such conditions. Moreover, these micron-sized grains within the cold regions of the dusty clumps can be covered by a water ice. The clumps can also harbour the micron-sized water ice grains. The depletion of oxygen by micron-sized grain covered by thick water ice and/or by micron-sized water ice grains in the diffuse ISM has been proposed as one of possible solutions to so-called `O crisis' \citep[see e.g.][]{2015MNRAS.454..569W}. To study the possibility of presence of large grains within the clumps, the model of \citet{Tsytovich_ea14} should be modified in order to take into account the grain size distribution.

The clumps with sizes below 0.1~au are optically thin for visible and UV radiation. The dust temperature within these clumps is similar to the dust temperature in the diffuse ISM region with the homogeneous dust distribution. The extinction in $V$-band of diffuse ISM region with such clumps for both a distant observer and observer within the region is practically the same as in the diffuse ISM region with homogeneous dust distribution. We conclude that the presence of small clumps with radii $<0.1$~au within diffuse ISM region does not have observed implications in terms of both the absolute value of extinction and, as we expect based on the results of \citet{2015A&A...584A.108S} for optically thin clumps, the extinction curve shape. We expect that the presence of such clumps in the medium can affect the chemical processes only via the mechanism considered by~\citet{Ivlev_ea18}.

\section*{Acknowledgements}

The authors are thankful to Prof. Paola Caselli and Dr. Yaroslav Pavlyuchenkov for critical reading of the manuscript and suggestions on its improvement. ABO, AIV, AVI and VAS were supported by the Russian Science Foundation via the project 18-12-00351. AIV and VAS are the members of the Max Planck Partner Group at the Ural Federal University. Monte-Carlo simulations were made by SYP. SYP acknowledges support by the Ministry of Science and Education, FEUZ-2020-0030.




\bibliographystyle{mnras}
\bibliography{mnras_template}




\section*{Appendix}

\subsection*{Effects of non-uniform dust density distribution in clumps}
\label{sec:nonuni}

To quantify effects of non-uniform density distribution on the extinction we fitted with fourth-degree polynomial, $P(r)$, the dust density distribution obtained by \citet{Tsytovich_ea14} for positively charged grains and normalized ion density of 0.5, which corresponds to maximum variations of the dust within clumps. This fit then has been used to calculate $\left<e^{-2\tau_{\rm{c}}} \right>$ based on equation~(35b) from~\citet{Conway_2018} as follows:

\begin{align}
\left<e^{-2\tau_{\rm{c}}} \right> &=  2\int\limits_0^1\exp \left[ -2\tau_{\rm{c}} \left( \int\limits_0^1 P(l)dl \right)^{-1} \right. \times \nonumber\\ & \times \left. \int\limits_0^{\sqrt{1-\rho^2}} P\left( 1+l^2-2l\sqrt{1-\rho^2} \right) dl \right] \rho d\rho  \,, \label{diff:nonuniexp}
\end{align}
where $\tau_{\rm{c}} \left( \int\limits_0^1 P(l)dl \right)^{-1}$ is a factor which normalizes non-uniform clump density so that optical to the clump centre equals to one obtained with the uniform dust density distribution in clump using equation~\eqref{model:tauc}. The maximum difference of $\left<e^{-2\tau_{\rm{c}}} \right>$ and, therefore, in extinction between the model with uniform and non-uniform clump density is of 30 per cent corresponding to largest clumps. Such a difference is lower than the difference in extinction between the clumpy and homogeneous models for large clumps. For the clumps with radii $<0.1$ au this difference does not exceed 2 per cent.

\begin{figure}
\centering
\includegraphics[width=0.95\linewidth]{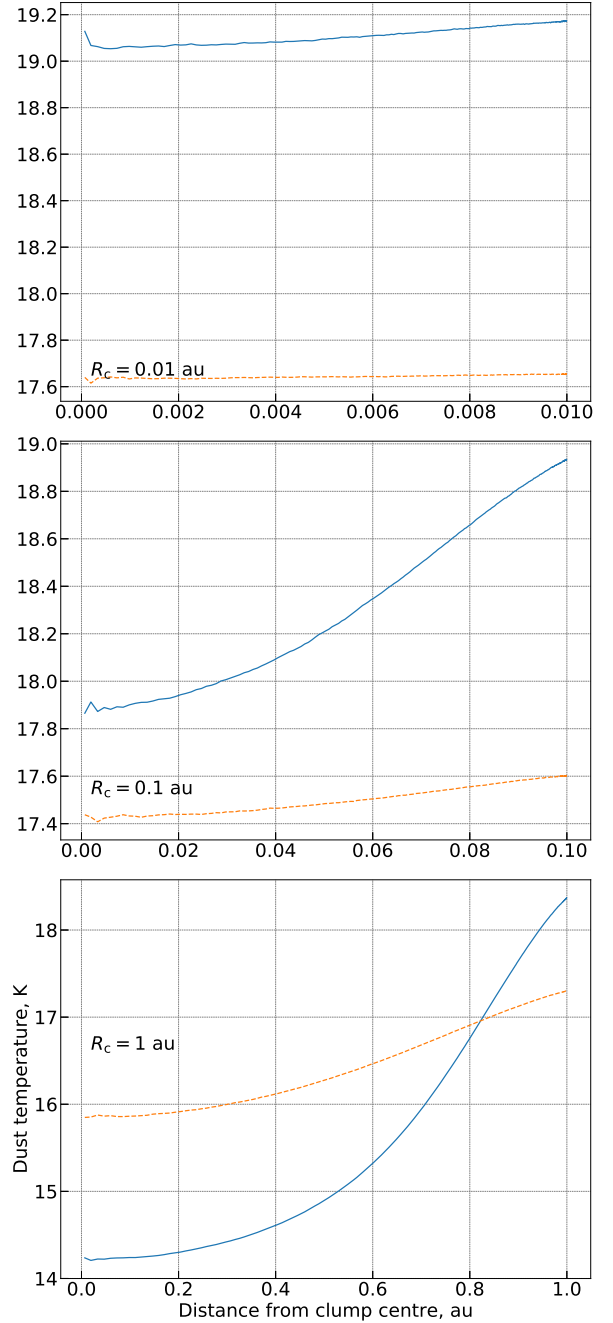}

\caption{Dust temperature profiles within clumps with different radius and non-uniform dust density distribution within clumps. The calculations were performed with unattenuated external starlight. Blue solid line: dust temperature calculated assuming that all dust grains have a radius of $a=0.03$~$\mu$m. Orange dashed line: dust temperature calculated with dust grain sizes distribution and optical properties from~\citet{2001ApJ...548..296W}.}\label{model:tempnonuni:fig}
\end{figure}

The dust temperatures calculated with non-uniform dust density within clumps (see Figure~\ref{model:tempnonuni:fig}) in the same manner as in Section~\ref{sec:td} do not differ from those obtained for the uniform dust density distribution within clumps by more than 0.5~K. Again, this difference is lower than the one between the clumpy and homogeneous models.

From these estimates, it follows that variations of the dust density distribution within clumps do not alter our conclusions on the optical properties and dust temperature in clumpy medium.

\subsection*{Grain and gas diffusion time}

The characteristic time $\tau_{\rm diff}$ for a dust grain to diffuse from the centre of a clump of radius $R_{\mathrm{c}}$ to its border can be estimated from the relation
\begin{align}
R_{\mathrm{c}}^{2}=6\,D\,\tau_{\mathrm{diff}} \,, \label{diff:5}
\end{align}
Here, $D$ is the self-diffusion coefficient of a grain. The value of $D$ is obtained from the expression of the self-diffusion coefficient of a heavy particle in a light gas (see e.g. equation (14.2.1) in~\citet{1970mtnu.book.....C}; \citet{1981ll10}):

\begin{align}
D=\frac{3}{8\sqrt{2\pi}\,a^{2}\,n_p}\,\sqrt{\frac{k_{\rm B}T}{\mu\,m_{\rm H}}}\,, \label{diff:grain}
\end{align}
where $a$~--- dust grain radius; $\mu$~--- molecular mass; $m_{\mathrm{H}}$~--- hydrogen atom mass; $T$~--- gas particles temperature within a clump which assumed to be equal to the dust temperature.

Then, for $n_{p}=10^{2}$~cm$^{-3}$, $T=20$~K, $a=0.1\times 10^{-4}$~cm, $\mu=1.5$, grain diffusion time from the centre of the clump to its edge is $\tau_{\mathrm{diff}}\sim 24$~kyr and 2.4~Myr for $R_{\mathrm{c}}=0.1$ and 1~au, respectively.

The time for gas particles to diffuse across a whole clump of radius $R_{\mathrm{c}}$, $\tau_{\rm{g}}$, can be estimated as ~\citep{1970mtnu.book.....C}:

\begin{align}
\tau_{\rm{g}} = \frac{1}{6} \left(2R_{\rm{c}}\right)^2 \frac{8}{3} \sigma n_{\rm{H}}\,\sqrt{\frac{2\pi \mu\,m_{\rm H}}{k_{\rm B}T}}\,, \label{diff:gas}
\end{align}
where $\sigma \sim 10^{-15}$ cm$^{2}$ is the gas-kinetic cross section of H atoms mutual collisions, and $n_{\rm{H}}$ is the gas density within a clump \citep[see][]{Ivlev_ea18}. The gas density within a clump is the ambient gas density of 100 cm$^{-3}$ enhanced by a factor of $\simeq T_{\rm{gas}}/T$, where $T_{\rm{gas}}$ is the gas temperature in the medium between clumps \citep[see][]{Ivlev_ea18}. Assuming $T=20$ and $T_{\rm{gas}}=100$~K, one can obtain that $\tau_{\rm{g}}=5$ and 480 yr for $R_{\mathrm{c}}=0.1$ and 1~au, respectively.

\subsection*{Extinction in the clumpy ISM region model}

Here in Table~\ref{model:tabavcmsym}, for convenience, we present numerical values of extinction in the clumpy model for some clump radii and for $A_{V}^{\rm{obs}}$ and region geometry considered in this study.

\begin{table}
  \caption{Optical properties of diffuse ISM region with clumps for different region geometries, clump radii and $A_{V}^{\mathrm{obs}}$.}\label{model:tabavcmsym}  \centering
  \begin{tabular}{rrr}
  \hline
  $R_{\mathrm{c}}$, au & $A_{V}^{\mathrm{cm}}$ ($A_{V}^{\rm{obs}}=0.3^m)$ & $A_{V}^{\mathrm{cm}}$ ($A_{V}^{\rm{obs}}=1.086^m)$ \\
  \hline
  \multicolumn{3}{c}{Spherical region with radius $L/2$} \\
  \hline
  0.1 & 0.14 & 0.51 \\
  0.5 & 0.11 & 0.41 \\
  1 & 0.09 & 0.33\\
  \hline
  \multicolumn{3}{c}{Infinite slab with thickness $L$} \\
  \hline
  0.1 & 0.43 & 1.16 \\
  0.5 & 0.37 & 0.99\\
  1 & 0.3 & 0.82\\
  \hline
  \multicolumn{3}{c}{Infinite cylinder with radius $L/2$} \\
  \hline
  0.1 & 0.21 & 0.70 \\
  0.5 & 0.17 & 0.57 \\
  1 & 0.13 & 0.46\\
  \hline
  \end{tabular}
\end{table}


\bsp	
\label{lastpage}
\end{document}